\documentclass[doublecol]{epl2}
\usepackage{latexsym}
\usepackage{graphicx}

\title{Is the nature of magnetic order in copper-oxides and in
 iron-pnictides different? }
\author{Efstratios Manousakis\inst{1,2} \and Jun Ren\inst{3} 
\and Sheng Meng\inst{3} \and Efthimios Kaxiras\inst{3}}
\shortauthor{E. Manousakis \etal}
\institute{
\inst{1}Department of Physics and MARTECH,
Florida State University,
Tallahassee, FL 32306-4350, USA\\
\inst{2}Department of  Physics, University of Athens,
Panepistimioupolis, Zografos, 157 84 Athens, Greece\\
\inst{3}Department of Physics and School of Engineering and
Applied Sciences, Harvard University,
Cambridge, MA 02138, USA
}
\abstract{We  use the  results  of  first-principles electronic  structure
calculations  and a  strong coupling  perturbation  approach, together
with general  theoretical arguments, to illustrate  the differences in
super-exchange    interactions   between    the    copper-oxides   and
iron-pnictides. We  show that  the two magnetic  ground states  can be
understood   in  a   simple   manner  within   the  same   theoretical
foundation. Contrary to  the emerging view that magnetic  order in the
iron-pnictides is of itinerant nature, we argue that the observed
magnetic  moment is  small because  of frustration  introduced  by the
electrons of the Fe orbitals as they compete to impose their preferred
magnetic ordering.}
\begin{document}


\maketitle

\begin{figure}[htp]
\begin{center}
\includegraphics[width=3.25 in]{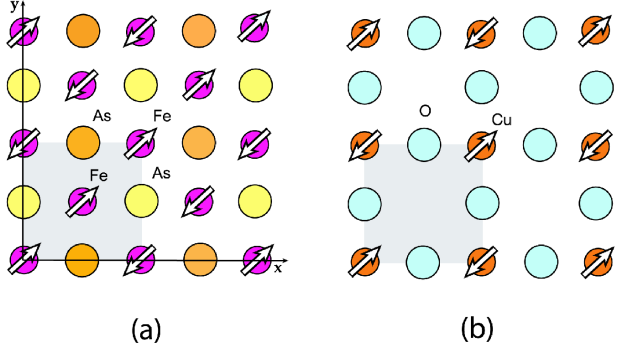}
\caption{ (a) The spin-density-wave order (columnar antiferromagnetic)
  as observed  by neutron diffraction.  The Fe  magnetic moments along
  the (1,1) direction are aligned, while two nearest neighbouring such
  chains   are  antiferromagnetically   aligned.   (b)  The   familiar
  antiferromagnetic ordering of the  cuprous oxides. The shaded square
  denotes the unit  cell.  With the mapping $Fe\to Cu$, $As\to O$, 
the $CuO_2$ unit  cell differs from the $Fe_2As_2$ one by a magnetic 
atom missing from the
  centre.}\label{fig1}
\end{center}
\end{figure}

The copper-oxide  layers  present  in the  high-Tc
superconducting   families   are   turned  into   superconductors   by
introducing dopants that create  electrons or holes in these otherwise
antiferromagnetic   (AF)   insulating   layers\cite{Vaknin}.  
The   proximity   of
antiferromagnetism to  superconductivity has  led to the  general view
that  this  form  of  magnetic  order is  intimately  related  to  the
mechanism of  superconductivity in  these materials\cite{RMP}. 
In  the recently
discovered  iron-pnictide   based  
superconductors\cite{Kamihara1,Kamihara2,Ren,Chen1,Chen2,Wen},  
which  exhibit
superconductivity  at relatively  high-Tc, the  copper-oxide  layer is
replaced  by an iron-pnictide  layer. Interest  in the  new materials,
reminiscent of that  seen when the cuprates were  discovered more than
two decades  ago, is due to  the fact that  many unsuccessful attempts
were made  to replace the  copper-oxide layer in high-Tc  materials to
facilitate  practical  applications.  The  parent compounds  of  these
iron-pnictide materials, like  the copper-oxide parent materials, show
a  spin-density-wave  order\cite{Lacruz,Klauss} illustrated  in  
Fig.~\ref{fig1}(a).  Unlike  the
copper-oxides, the parent compounds in the iron-pnictides, such as the
pure  LaOFeAs,   are  metallic,  but  are   magnetically  ordered  and
non-superconducting  and they  become superconductors  by  doping with
electrons  or  holes.   The  copper-oxide parent  compounds  are  well
described         as        spin-1/2         Heisenberg        quantum
antiferromagnets\cite{RMP}.   Furthermore,   it   is   widely   believed   that
superconductivity in copper-oxides arises  when, by doping the quantum
antiferromagnet,  the carriers  (holes or  electrons) form  a strongly
correlated Anderson-Mott  type system with  the spin-spin correlations
playing a fundamental role  in the superconductivity mechanism.  After
the  discovery  of  the  iron-pnictide superconductors,  there  is  an
emerging view  that the magnetism  in these compounds is  of itinerant
type\cite{Singh,Cao,Raghu,Chubukov} and that  these systems are different 
from  the cuprates and
in  fact weakly  correlated. This  is  an important  issue to  settle,
because any  further theoretical analysis of other  properties of this
new family of materials, including the still unknown superconductivity
mechanism, depends  on it.  

 Here we  focus on this  issue and  seek a
broader framework  to reconcile the different forms  of magnetic order
and to explain  the magnetic properties in both  families of materials
with the same approach.  Our  arguments are inspired by the results of
Ref.~\cite{Kaxiras},  where  first-principles electronic  structure  
calculations
based on density  functional theory (DFT) were combined  with a strong
coupling expansion to obtain an effective low-energy Hamiltonian which
describes  the  electrons occupying  the  five  Fe  orbitals.  In  the
present  work, we  address the  nature of  the magnetic  order  in the
iron-pnictide  and  the  copper-oxide  based materials  using  a  more
general scheme,  which is correct  independently of the  conditions of
validity and the  details of the calculation presented  in Ref. 15. We
show that the origin and nature  of magnetism in these two families of
materials is the  same and both families should  be treated within the
same  theoretical   foundation.  Specifically,  if   Mott  physics  is
operative  in the  copper-oxides it  should also  be operative  in the
iron-pnictides, and  if the magnetism  in the copper-oxides is  of the
Anderson-Mott  type it  should  be  of the  same-type  in these  newly
discovered  materials.   

First,  we  discuss  the  difference  in  the
structure and  magnetic order between the iron-pnictide  layer and the
copper-oxide layer as shown in Fig.~\ref{fig1}. By removing the “extra” magnetic
ion  (i.e.,  Fe) from  the  centre  of the  square  unit  cell of  the
iron-pnictide  layer  we  obtain  not  only the  same  structure,  but
identical  magnetic  order  with  the  copper-oxide  layer.  The  only
important  difference, which  will  be addressed  below,  is that  the
magnitude  of   the  observed  magnetic   moment\cite{Lacruz}  per  
Fe   site  is
significantly  reduced from its  calculated 
value\cite{Singh,Yildirim,Ma,Haule}.  

\begin{figure}[htp]
\begin{center}
\includegraphics[width=3.25 in]{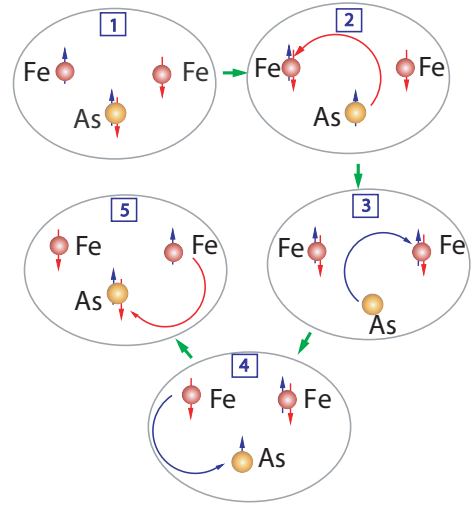}
\caption{  Super-exchange between two  nearest neighbour (NN)  or next
nearest   neighbour   (NNN)    Fe   atoms   through   electron-hopping
(hybridization) process between any Fe d-orbital and any p or s orbital 
of an intervening As atom.   One of the As spin-up  electrons (shown as red)
hops to a  singly occupied Fe d-orbital (process  2), then the spin-down
electron (shown as blue) from the same  As orbital hops to a NN or NNN
Fe d-orbital (process  3). With two successive hops  (processes 4 and 5)
the  other electrons  that were  initially occupying  the same  two Fe
d states, hop to the same As site, restoring its doubly occupied status.
The initial configuration 1 differs  from the final configuration 5 by
the spin orientation  on the two Fe orbitals.   Such processes lead to
AF  spin-spin interaction  between  electron spins  occupying any  two
d orbitals belonging to NN or NNN Fe atoms.}\label{fig2}
\end{center}
\end{figure}

 Fig.~\ref{fig2} schematically  illustrates the  qualitative origin  
of  the well-known
Anderson super-exchange interaction. This  type of processes give rise
to a  super-exchange contribution to the  spin-exchange interaction,
$J^{(1)}_{\nu}$ , between  two electrons  occupying  
two nearest  neighbour  (NN), or  $J^{(2)}_{\nu}$ ,
between  two next  nearest  neighbour  (NNN) Fe  orbitals  of a  given
flavour (where  $\nu$ labels the five Fe or  Cu d-orbitals: $x^2-y^2$,$xz$,$yz$,
 or $z^2$ ). The
effective spin-spin  interaction Hamiltonian for  a particular flavour
$\nu$ is:
\begin{eqnarray}
\hat H_{\nu} = J^{(1)}_{\nu} \sum_{<ij>} {\bf S}_{i,\nu} \cdot {\bf S}_{j,\nu} +
J^{(2)}_{\nu} \sum_{<<ij>>} {\bf S}_{i,\nu} \cdot {\bf S}_{j,\nu},
\label{Hamiltonian}
\end{eqnarray}
where the ${\bf S}_i$ is  a three component vector of a spin-1/2 quantum
spin  operator and $<ij>$ and $<<ij>>$ stand  for NN  and NNN  
pairs of  magnetic ion
sites.  Here we have assumed  that the particular Fe d-orbital is singly
occupied. The case in which an orbital is occupied by two electrons is
discussed  separately. Other  models in  the literature,  as  in 
Refs.\cite{Xu,Yildirim,Ma}, are similar 
to ours with the very important difference that our
Hamiltonian,  Eq.\ref{Hamiltonian}, 
treats  differently  the  various Fe  d-orbitals
labelled by $\nu$.  The condition for quantum- N\'eel (checkerboard) ordering
(shown  in Fig.  5(b))  is
\begin{eqnarray}
J^{(1)}_{\nu} > 2 J^{(2)}_{\nu};
\label{condition}
\end{eqnarray}  
if the  coupling  between spins  of
electrons occupying  the orbital $\nu$ do  not satisfy the  above condition,
the  SDW order  of Fig.\ref{fig1}(a)  is energetically  favourable  against the
checkerboard  order  (Fig.~\ref{fig5}(b)).   

\begin{figure}[htp]
\begin{center}
\includegraphics[width=3.25 in]{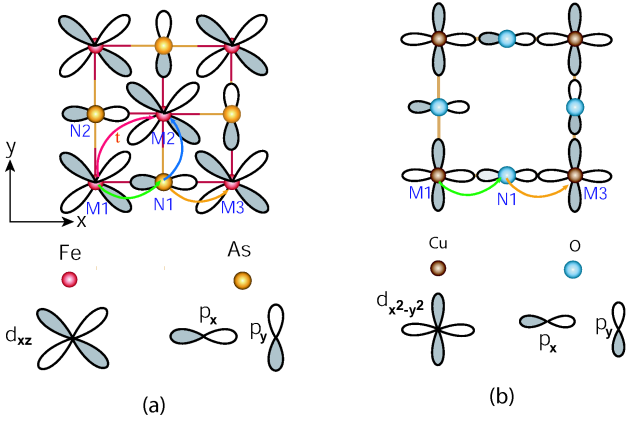}
\caption{(a)  An  example  of  the paths  that  contribute  to  the
spin-exchange interactions $J^{(1)}_{\nu}$ and $J^{(2)}_{\nu}$ 
between  NN and NNN  Fe atoms  for $\nu=d_{xz}$ and
through the $p_x$ or $p_y$ orbital of the intervening non-magnetic  
As atoms. (i) Contributions to  $J^{(1)}_{\nu}$ 
of the  NN Fe  atoms indicated as  M1 and  M2: First
there is  a contribution  from direct hopping  t between  the magnetic
ions  as shown by  the red  path. Second  there is  the super-exchange
through the $p_x$ or $p_y$  orbital (only the $p_x$ is drawn) 
of the intervening As atom
N1  or  N2.  One of  these  is  shown  by  the green-blue  path.  (ii)
Contributions to $J^{(2)}_{\nu}$ 
for  the NNN magnetic atoms M1 and  M3 through the $p_x$ or $p_y$ 
orbital of  the intervening As atom  N1 shown as  a green-yellow path.
(b) The  corresponding situation for the copper-oxide  layer: There is
no  corresponding  M2  magnetic  atom and  the  spin-spin  interaction
between the  NN Cu atoms  M1 and M3  for this case occurs  through the
hybridization of the $d_{x^2-y^2}$ orbital of the Cu atom M1 or  M3 and 
the $p_x$ orbital of the intervening O atom N1.}\label{fig3}
\end{center}
\end{figure}

By fitting  the  first-principles
electronic  structure  results  to  a  tight-binding  model,  we  have
established that, with the exception  of the $d_{x^2-y^2}$ orbital, for 
any given Fe
d-orbital, there is only one orbital  of the intervening As (of s or 
p mixed
character) with which the hybridization amplitude is most significant.
Therefore, the  most significant super-exchange  processes involve the
same intervening As orbital for  both of the exchanged electrons.  The
super-exchange  process  illustrated  in  Fig.~\ref{fig2}  requires  four  steps
because  two  electrons, each  occupying  one of  the  two  NN or  NNN
magnetic (Fe)  ions, need to  move through hopping to  the intervening
non-magnetic ion (As) and, thus,  each electron has to make two steps.
To obtain  the total spin-spin  interactions we need to  consider also
the direct hopping  of electrons between these two  neighbouring Fe or
Cu  d-orbitals. The  results of  Ref.~\cite{Kaxiras}  show that  
the  direct hopping
matrix element between two NN Fe d-orbitals, indicated as $t$ in 
Fig.~\ref{fig3}, is
smaller but not  negligible. In the case of  copper-oxides this direct
process is non-applicable because two NN Cu atoms correspond to NNN Fe
atoms in the Fe-pnictides.

\begin{figure}[htp]
\begin{center}
\includegraphics[width=3.25 in]{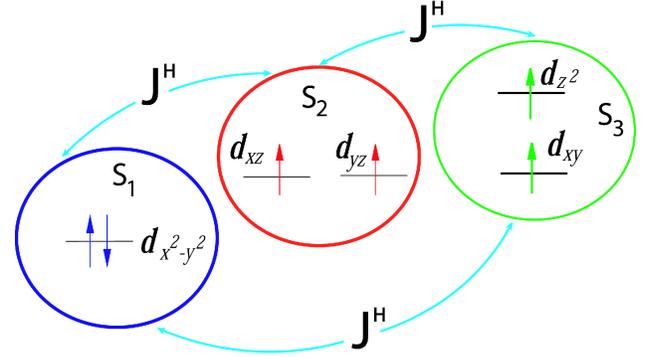}
\caption {The Hilbert space of  the five Fe d-orbitals  is divided into
three subspaces  ($S_1$ spanned by $d_{x^2-y^2}$, $S_2$ spanned  by 
$d_{xz},d_{yz}$, and $S_3$ spanned  by $d_{xy},d_{z^2}$ ). These
subspaces  are  coupled  through   the  Hund's  rule  coupling $J^H$ which is
indicated by the blue arrows.}\label{fig4}
\end{center}
\end{figure}

\begin{figure}[htp]
\begin{center}
\includegraphics[width=3.25 in]{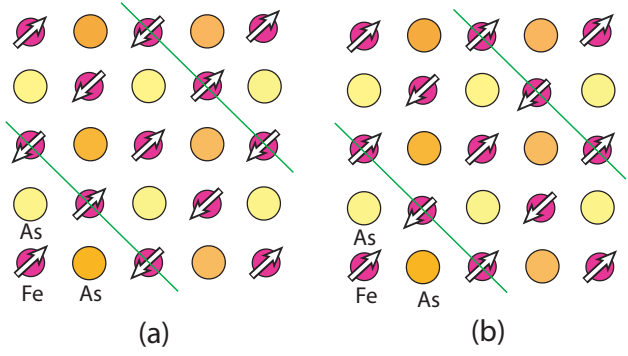}
\caption{(a) Illustration of the  observed columnar AF  order in the
undoped  iron-pnictides.  This  state  is  strongly  favoured  by  the
$S_2$ subspace (spanned by $d_{xz},d_{yz}$).  (b) Illustration of the 
familiar checkerboard
N\'eel ordering which is preferred by  the $S_1$ subspace (spanned 
by $d_{x^2-y^2}$) and by
the $S_3$ subspace (spanned by $d_{xy},d_{z^2}$). Notice that the magnetic 
moments of the Fe
atoms along the green  lines are frustrated because their orientations
in case (a) are opposite to their orientations in case (b).}\label{fig5}
\end{center}
\end{figure}

Fig.~\ref{fig3} illustrates paths that give rise to 
the  most   significant  contribution  to $J^{(1)}_{\nu}$ and $J^{(2)}_{\nu}$, 
  and  compares  the
super-exchange  interactions between the  iron-pnictide layer  and the
copper-oxide layer  through the intervening non-magnetic As  or O atom
respectively. For example,  for iron-pnictides, $J^{(1)}_{xz}$has 
two contributions:
(i) one from the direct hopping $t_{xz}$ between two NN Fe 
$d_{xz}$ orbitals, (shown as
the red path in Fig.~\ref{fig3}), and (ii) a super-exchange contribution due to
processes such as the one  illustrated by the combined green-blue path
in Fig.~\ref{fig3}. The hybridization $V^x_{xz}$ of the $d_{xz}$ orbital of the 
Fe atom M1 and the $p_x$
orbital  of  the  As atom  N1  is  very  large  ($\sim 1.4$ eV),  
while  its hybridization $V^y_{xz}$ with the $p_y$ orbital of the same 
atom N1 is negligible ($\sim 0.1$
eV). The reverse is true for  the hybridization of the same Fe d-orbital
with the $p_x$ and $p_y$ orbitals  of atom  in position N2  because of 
 the {\it $90^{\circ}$ relative
orientation}.   The  contribution  to  $J^{(1)}_{xz}$ 
due  to  the  green-red  process
illustrated in  Fig.~\ref{fig4}  is proportional  to 
\begin{eqnarray}
J^{(1)}_{xz} \sim (V^x_{xz}V^y_{xz})^2.
\end{eqnarray}  
On the other hand, the contribution to  $J^{(2)}_{xz}$
due to the green-yellow process illustrated in Fig.~\ref{fig4}
is proportional to 
\begin{eqnarray}
J^{(2)}_{xz} \sim (V^x_{xz})^4.
\end{eqnarray}  
Therefore, $J^{(2)}_{xz}/J^{(1)}_{xz} \sim (V^y_{xz}/V^x_{xz})^2 \sim 0.01$ and 
the super-exchange contribution to  $J^{(1)}_{xz}$ is
negligible as compared to that of  $J^{(2)}_{xz}$ between NNN Fe atoms. 
Hence, we can
conclude that  in the subspace formed by $d_{xz}$ and $d_{yz}$,  $J^{(1)}_{xz}$
has contributions mainly
through   direct   hopping   while $J^{(2)}_{\nu}$ has  
 significant   super-exchange
contributions similar to the  copper-oxide case, as shown in 
Fig.~\ref{fig3}(b).

On the other hand, the  NN spin-spin interaction  $J^{(1)}_{\nu}$, involving 
the other
three  Fe d-orbitals  ($yz$,$xy$, or $z^2$ ), has  contributions  from super-exchange
processes  in  which the  intervening  As  orbital  is the $sp_z$ (a  linear
combination of the As $4s$ and $4p_z$). In these processes the magnitude of the
hybridization of  the $yz$,$xy$, or $z^2$ Fe d-orbitals  with the $sp_z$-As  
orbital {\it does not change}  with a  $90^{\circ}$ rotation  which 
necessarily  occurs  in the  red-green
path.  Therefore, for  these three  orbitals,  $J^{(1)}_{\nu}$ is  
significantly larger
than $J^{(2)}_{\nu}$, which implies that  the condition (\ref{condition}) 
for checkerboard order is fulfilled.  

Fig.~\ref{fig4} illustrates the fact that the Hilbert space spanned
by all the Fe d-orbitals  can be separated into three subspaces: $S_1$ spanned
by $d_{x^2-y^2}$, $S_2$ spanned by  ($d_{xz},d_{yz}$), and 
$S_3$ spanned by ($d_{xy},d_{z^2}$).  These subspaces  are mainly
coupled through the Hund's rule coupling $J^H$ which tends to align the spins
of the unpaired d electrons in the  same Fe atom:
\begin{eqnarray}
\hat H &=& \sum _{\nu=1}^5 \hat { H}_{\nu} -
J^H \sum_{i,\nu\ne \nu'} {\bf S}_{i,\nu} \cdot {\bf S}_{i,\nu'}.
\end{eqnarray}
 The NN AF couplings 
$J^{(1)}_{\nu}$ in
the subspaces $S_1$ and $S_3$ are greater than  the NNN AF couplings 
$J^{(2)}_{\nu}$; however, in
the  subspace $S_2$ the  NNN AF  coupling $J^{(2)}_{xz}$ (and $J^{(2)}_{yz}$)  
is  large, that  is,  the
condition (\ref{condition}) is not satisfied for the arguments presented 
above. As a result the  subspace $S_2$ alone prefers the observed  
SDW state illustrated
in   Fig.~\ref{fig5}(a),  while   the   other   two   subspaces  prefer  
 the checkerboard-type  quantum-N\'eel order  shown in  Fig.~\ref{fig5}(b).  
These two
competing  orders   create  a  magnetic   frustration  illustrated  in
Fig.~\ref{fig5}. Namely,  in the two ordered states illustrated  in 
Fig.~\ref{fig5}, the
magnetic  moments  of  the  electrons  on the  iron  atoms  along  the
alternating  green lines  in  Fig.~\ref{fig5} are  opposite.  
The Hund's  rule
coupling, which is significantly stronger that the AF coupling between
NN and NNN atoms, forces the electrons on these different subspaces to
choose  a common spin  orientation.  However,  any such  global choice
within the  same atom will minimize the energy in one subspace and 
at  the same time
frustrate the other subspaces as discussed in the caption of Fig.~\ref{fig5}. 
In the  case  of the  undoped  material  where  there are  six  electrons
occupying the five Fe d-orbitals, the $S_1$ subspace has zero spin; hence, 
the $S_2$ subspace competes with the $S_3$ subspace  
(which prefers the checkerboard AF
order)  only and  because $J^{(2)}_{xz}$ is  larger  than the  
AF couplings  in $S_3$,  it
imposes the observed SDW state.  The local magnetic moment is expected
to be  small due to this  frustration. If we neglect  the influence of
the  other subspaces,  and  we  restrict ourselves  to $S_2$, the  maximum
expected order should  be less than $2\mu_B$  because of  the reduction of the
magnitude  of the  order parameter  from  its classical  value due  to
zero-point spin fluctuations. If we  turn on the interaction with the 
other subspaces, this has a frustrating effect of the magnetic moment.
Therefore, the present analysis gives a natural explanation for 
the observed small magnitude
of the magnetic moment\cite{Lacruz}, while  the calculations based on 
the itinerant
picture\cite{Singh,Cao,Yildirim,Ma,Haule} produce  values  for  the  
Fe  magnetic  moment greater than $2\mu_B$. 

 From this  analysis, we have shown that the different
ordering  in the  iron-pnictides and  copper-oxides can  be understood
within the same theoretical foundation. The observed small Fe magnetic
moment  arises from  the fact  that only  the electrons  occupying the
subspace spanned by $d_{xz},d_{yz}$ prefer the observed SDW order  
(Fig.~\ref{fig5}(a)), while
the  electrons  occupying  the   other  three  Fe  orbitals  prefer  a
checkerboard ordering  (Fig.~\ref{fig5}(b)), which create  large 
zero-point spin fluctuations.   Including  these frustrating  effects,  
we expect  a
significantly   reduced  local  moment   in  agreement   with  neutron
scattering  experiments.  



\acknowledgements 

We have benefited from discussions with C. Xu, E. Demler, S. Sachdev and 
B.I. Halperin.

\end{document}